# CLAIRE: Combining Sets, Search and Rules to Better Express Algorithms

*Yves Caseau, François-Xavier Josset, François Laburthe*

Bouygues SA – e-lab
1 avenue Eugène Freyssinet
78061 St-Quentin en Yvelines cedex, FRANCE
{ycs,fxjosset,flaburth}@challenger.bouygues.fr

**Abstract**. This paper presents a programming language that includes paradigms that are usually associated with declarative languages, such as sets, rules and search, into an imperative (functional) language. Although these paradigms are separately well-known and available under various programming environments, the originality of the CLAIRE language comes from the tight integration, which yields interesting run-time performances, and from the richness of this combination, which yields new ways to express complex algorithmic patterns with few elegant lines. To achieve the opposite goals of a high abstraction level (conciseness and readability) and run-time performance (CLAIRE is used as a C++ pre-processor), we have developed two kinds of compiler technology. First, a pattern pre-processor handles iterations over both concrete and abstract sets (data types and program fragments), in a completely user-extensible manner. Second, an inference compiler transforms a set of logical rules into a set of functions (demons that are used through procedural attachment).

## 1. Introduction

The CLAIRE language [CL96a] is born from a research project that started 6 years ago which goal was to study the relationship between combinatorial optimization and constraints. We decided to evaluate the competitive advantages and reciprocal benefits of combinatorial optimization (CO) and constraint programming (CP) techniques, and then to combine them into hybrid algorithms. We followed a systematic approach and looked at problems such as Resource Allocation, Matching, Scheduling (job-shop, cumulative), Routing (TSP, VRP, VRPTW) or Time-Tabling. Most real industrial optimization problems are a combination of these more classical problems, thus another goal of this project was to identify a set of "building blocks" for optimization applications. For each of these problems, we have produced a library of hybrid algorithms.

The combination of hybrid algorithms can only be made at the source level, using a "white box" approach. The complexity and the inter-dependence between the various algorithms for propagation, computing lower bounds, branching, etc. is such that an approach based on components (black boxes) does not seem realistic to us (today) if one wants to take advantage of all the various techniques provided by the CO field. Thus, we have found the need for a language to express our different algorithms that would be:

- *simple and readable*, for many reasons: because reuse of algorithms requires a readable expression of the ideas, but also because CLAIRE was part of a research and teaching project and we needed an "executable pseudo-code" to demonstrate algorithms;
- *multi-paradigm* (supporting logic, imperative and functional programming styles), with few simple and well-understood concepts, such as objects, functions, rules and versioning for building search trees;
- *compiled efficiently* so that we could compare our algorithms with state-of-the-art approaches from teams in the CO community who use C or FORTRAN as their implementation language. We decided to build CLAIRE as a C++ preprocessor that would generate hand-quality C++ code (CLAIRE has also been ported to Java).

The paper is organized as follows. The next section is a presentation of CLAIRE at a glance. CLAIRE is an object-oriented functional language with parametric polymorphism. It is also a modeling language that uses high-level abstractions such as sets or relations. Last, it includes an inference engine for object-based rules and a few tools for building tree search algorithms. The third section presents the notion of "set programming" in CLAIRE, which is derived from a set of techniques that make the use of sets as a programming paradigm both natural and efficient. The fourth section shows how logical rules are compiled into functions. This is based on previous work developed for the LAURE language [Cas94] (one of CLAIRE's ancestors), so we focus on the application of the technique to the CLAIRE logic and the impressive performance that we now obtain (since CLAIRE functions are compiled into C++ code with virtually no overhead). The last section shows the application of CLAIRE to writing optimization algorithms. We show how we can use the built-in features to write tree search algorithms, and how the strategy of source-to-source optimization introduces new opportunities for optimization through the notion of composition polymorphism.

## 2. CLAIRE at a Glance

We simply give an overview of the most salient features of the language. A complete description, including a formal syntax for CLAIRE may be found in [CL96a].

### 2.1 A Simple Object-Oriented Data Model

We first briefly describe the CLAIRE object-oriented data model. Classes are defined with a list of typed slots, and only single inheritance is supported. Each class is a first-class object of the system and is accessible at run-time (an optional list of instances is stored for each class). The `<:` operator is used to define a class (because subclasses are subtypes in CLAIRE); it consists of appending a list of new slots to the slots of the parent class. Slots support default values and unknown values, i.e., there is a special `unknown` object that represents the absence of value. The following two examples illustrate class definition; the second one shows a parameterized class (the usual `stack` class, parameterized by the type of its elements).

```
Point <: object(x:integer = 0, y:integer = 0)
Stack<of> <: object(of:type, content:list = nil)
```

The second example is slightly surprising, compared to other approaches such as template classes, because it implicitly rely on the fact that types are first-class objects (hence the `type`). Thus we attach to each stack the type that must contain the objects that are pushed in the stack through the slot *of*. The responsibility of checking that objects in the *content* of the stack belong to this type is left to the programmer. On the other hand, parameterized signatures, as we shall see later, can express this dependency to support the strong typing of stack operations.

The CLAIRE data type complements the simplicity of the class hierarchy. A type is either a class, a constant set of objects, the union or the intersection of two types, a parameterized class, an integer interval or a typed subset of `list`, `array` or `set`. A parameterized class is the subset of the class such that the parameter belongs to a given type. For instance, `Stack[of:{integer,float}]` is the set of stacks whose type parameter is `integer` or `float`. A stack from this parameterized type will contains either integers (if its *of* parameter is `integer`) or floats, but not both. A stack that contains either would be represented by the parameterized type `Stack[of:(integer ∪ float)]`

There are three kinds of primitive collections: arrays (constant length and ordered), lists (dynamic length, ordered) and sets (dynamic length, no order and no duplicates). Primitive collection can either be typed, with a read-only type attribute `of`, similar to the `stack` example, or un-typed and immutable. This means that only typed lists or sets support updates, which makes sense since we need a type to guarantee the type-safety of updates. Immutable lists (or sets) receive a dynamic type when they are created.

Consequently, both list and set types come in two flavors: either strongly typed (`list<t>`, *à la* C++), or supporting inclusion polymorphism (`list[t]`), i.e., $list[t_1] \leqslant list[t_2] \Leftrightarrow t_1 \leqslant t_2$. The type `list<t>` contains typed lists l whose type parameter `of(l)` is precisely `t`. A `list<t>` object is a list that contains objects that belong to `t`, including instances of subtypes of `t`.

On the other hand, the type `list[t]` contains both typed lists l whose type parameter `of(l)` is a subtype of `t`, as well as immutable lists whose member belong to `t`. This applies equally to `set<t>` and `set[t]`.

The distinction between these two kinds of list/set types is a key feature in CLAIRE. Since `list<t>` ≼ `list[t]`, we can write generic list methods, that apply both to mutable and dynamic, immutable lists. The type `list[t]` is interesting because of inclusion polymorphism, and because read operations on lists of type `list[t]` can be statically type-checked. On the other hand, `list<t>` is monomorphic ($list<t_1> \leqslant list<t_2> \Rightarrow t_1 = t_2$), but both read and write operations can be statically type-checked [Cas95]. Let us briefly illustrate this on an example. Consider three classes A ≤ B ≤ C and the following methods:

```
foo(x:list[C], y:B) : void -> x[1] := y
bar(x:list<B>, y:B) : void -> x[1] := y
```

Although B is a subclass of C, the insertion in the `foo` method into the list x is not safe, since x may be a `list<A>` object; on the other hand, `bar` is safe but less flexible. Thus, the method `foo` relies on dynamic typing at run-time to guarantee the safety of the update, whereas the method `bar` is compiled more efficiently.

In addition, the `tuple` type is used for constant, heterogeneous, typed "array-like" (i.e., indexed) containers; for instance, `tuple(integer,integer,float)` represents the set of tuples with exactly three elements: two integers and a float. For example, tuples are used in CLAIRE to return multiple values from a function.

We can summarize the richness of type expressions with the following grammar (as usual, `t[]` is the type for arrays of `t` (for instance, `t = integer`) and `..` is used to define integer intervals):

```
<type> ::=    <class>                  |  <class>[:<type> *]  |
              {<object> *}             |  (<integer> .. <integer>)       |
              (<type> ∪ <type>)        |  (<type> ∩ <type>)              |
              set⟨<type>⟩              |  list⟨<type>⟩                   |
              set[<type>]              |  list[<type>]                   |
              <type>[]                 |  tuple(<type> *)                |
              subtype[<type>]
```

Types are conceptually represented as expressions that denote sets of objects, and can be used everywhere in a program since they are reified. Types as objects are themselves typed with the `subtype[...]` type; for example, `subtype[integer]` contains all types that are included in `integer`. This set-oriented view is supported by the fact that type subsumption is defined by the inclusion of the set extensions for all possible evolution of the object database. Types form an inclusion lattice with syntactical greater lower bound and subsumption (subtyping). The relationship between the type lattice and the powerset of the object domain is the basis for using an abstract interpretation scheme [CP93].

CLAIRE inherits the SMALLTALK[GR83] strategy of considering everything as an object, from a class. This implies that primitive types such as integer, floats, strings, Boolean are seen as primitive classes in CLAIRE.

## 2.2 Polymorphism and Advanced Typing

Methods in CLAIRE are simply overloaded functions, where each parameter contributes to the overloading (the so-called *multi-methods* [CL94]) and each parameter can be typed with *any* CLAIRE type. For instance, attaching a method to a class union is useful to palliate the absence of multiple inheritance. Suppose that a class A is a natural candidate for multiple inheritance with parents B and C. If we chose B as the unique parent of A, we can define the methods for C that we want to be inherited by A on the explicit union (A ∪ C). This is not equivalent because of the lack of extensibility (adding a new subclass to (A ∪ C) is only possible through the addition of a subclass to A or C). However, it is close enough for most practical cases.

Methods may be attached anywhere in the type lattice, not simply to the class hierarchy (more precisely, anywhere in the lattice of type Cartesian products). This makes CLAIRE closer to a language like Haskell than to C++. A method is called a *restriction* of the global function which itself is called a *property*.

It is important to notice that the class hierarchy is considered as a "law" given by the user (seen as set inclusion) and is not based on type substitution. As a consequence, there are no constraints imposed on the user when adding new methods. Covariance is enforced implicitly but not imposed explicitly (adding a new method on a subdomain may augment the range of the function on a larger domain). If we define:

```
foo(x:integer) : (0 .. 5) -> ...        // ... stands for "some expression"
foo(x:(1 .. 10)) : (5 .. 10) -> ...
```

the range of `foo` on integer is actually the union (0 .. 10). The range in the first definition is taken as a type restriction for this first definition, not for "foo on any integer" which is defined by the two restrictions. Covariance comes as no surprise since we are using an overloading approach, where the property is the "overloaded function" and the methods (i.e., the restrictions of the property for a given signature) are the "ordinary functions" using Castagna's terminology [Cas95].

The use of a type lattice does not prevent the occurrence of "inheritance conflicts" where two restrictions have conflicting definition on overlapping signature. If one domain is included in the other it is assumed to take precedence (such is the case for the foo example, so foo(3) uses the second definition). Otherwise, a warning is issued. The lattice structure guarantees that it is possible to solve the inheritance conflict with a proper, more specialized, definition, since the intersection of overlapping signatures can always be described with types.

Parameterized types can be used to produce parameterized signature. That means that we can use the value of a parameter, when it is a type, inside the signature, as in the following example (for each operation `op`, `x :op y` is syntactical sugar for `x := (x op y)`):

```
push(s:Stack<X>, y:X) : void -> s.content :add y
```

The use of the type variable X states that the second argument must belong to the type contained in the slot `of` from the first argument (the stack). Note that there is an explicit difference between `stack[of:{integer}]` and `stack[of:subtype[integer]]`. The first is the set of stacks with *range parameter* integer (exactly), whereas the second can be thought of as the stacks of integers (stacks whose *range parameter* is a subtype of integer). There is no parametric subtyping with the first (i.e. `stack[of:{integer}]` ≼ `stack[of:{number}]` is false) as noticed in [DGL+95]. However, there is parametric subtyping with the second (i.e. `stack[of:subtype[integer]]` ≼ `stack[of:subtype[number]]`). To alleviate notations, `stack<integer>` is syntactical sugar for `stack[of:{integer}]`

Another important feature is the fact that CLAIRE supports higher-order functions. Properties, that are function names, may be passed as arguments to a method and they can, in turn, be applied to functions. For instance, one can write:

```
exp(f:property, x:any, n:integer) : integer           // exponentiation
    -> if (n = 0)   f(x)
       else   f(exp(f, x, n - 1))
```

This example will not be type-checked by the compiler and CLAIRE will rely on dynamic typing. On the other hand, most methods are statically type-checked and the C++ code produced by the compiler is very close to hand-written code (on classical benchmarks for procedural languages, using CLAIRE is similar to using C++, as is shown in the appendix).

Last, second-order types can be attached to methods. A second-order type is a method annotation, defined by a lambda abstraction, which represents the relationship between the type of the input arguments and the result. More precisely, a second-order type is a function such that, if it is applied to any valid type tuple for the input arguments, its result is a valid type for the result of the original function. Let us consider the two following examples:

```
Identity(x:any) : type[x] -> x

top(s:Stack<x>) : type[x] -> last(s.content)
```

The first classical example states that `Identity` is its own second-order type. The second one asserts that the top of a `Stack` always belong to the type of the `stack` elements (`of` parameter). In the expression e that we introduce with the `type[e]` construct, we can use the types of the input variables directly through the variables themselves, we may also use the extra type variables that were introduced in a parameterized signature (as in the `top` example) and we may use the basic CLAIRE operators on types such as ∪ (type union), ∩ (type intersection) and some useful operations such as `member`. `member` is used in particular to describe the range of the enumeration method `set!`. This method returns a `set`, whose members belong to the input class c by definition. Thus, we know that they must belong to the type `member(x)` for any type x to which c belongs (If c is a type, `member(c)` is the minimal type that contains all possible members of c). This translates into the following definition:

```
set!(c:class) : type[set[member(c)]] -> instances(c)
```

In the same way that overloading becomes more powerful with a richer type system, we have introduced tables to represent binary relations. Tables extend traditional arrays by allowing the domain (and the range) to be any type. When the domain is not an integer interval (the usual case for arrays) it is implemented through hashing. The following examples illustrate CLAIRE tables:

```
square[x:(0 .. 20)] : integer := x * x

dist[x:(0 .. 10), y:(0 .. 10)] : float := 0.0

creator[c:class] : string := "who created that class"

color[s:{car, house, table}] : set<colors> := {}

married[t:tuple(person,person)] : tuple(date,place) ∪ {nil} := nil
```

The combination of objects, relations (with explicit and implicit inverses), heterogeneous sets and lists makes CLAIRE a good language for knowledge representation. For instance, entity-relationship modeling is straightforward.

### 2.3 Set and Logic Programming

Sets play an important role in CLAIRE. We saw that each type represents a set and may be used as such. A set is used through testing membership, iteration, or combining with another set and a set operation. Membership is tested using the operation ∈ (the ASCII representation is '%'). Iteration is performed with the `for x in S e(x)` control structure. We have also introduced two convenient expressions:

- `exists(x in S | P(x))` returns `true` if there exists an x in S such that `P(x)` is true, and `false` otherwise
- `some(x in S | P(x))` returns an x from S such that `P(x)` is true if one exists, and `unknown` else (as in ALMA [AS97]).

For instance (refer to [CL96a] for further information), we may write:

```
x ∈ (string ∪ list)

for y in (1 .. 10) ∪ (20 .. 30)
     (if exists(c in (1 .. 10) | ok?(c + y))  choose(y))

when x := some(t in TASK | completed?(t)) in
     register(x)                              // t was found
else error("no completed task was found")
```

Although primitive data types (such as string or list) may be considered as sets, they are seen as infinite sets, thus trying to enumerate them will yield an error.

New sets can be formed through selection (e.g., `{x in S | P(x)}`) or set image (e.g., `{f(x) | x in S}`) [SDD+86]. When duplicates should not be removed, a list image can be formed with `list{f(x) | x in S}`. Using a straightforward syntax leads to straightforward examples:

```
{x in Person | x.age > 18}
exists(p in {x.father | x in (man U woman)} | p.age < 20)
list{x.age | x in Person}
for x in {f(y) | y in (1 .. 10)}  print(x)
sum(list{y.salary | y in {y in Person | y.dept = sales}})
```

In addition to sets, CLAIRE uses a fragment of an object-oriented logic to support inference rules associated to events, which will be presented in Section 4. This logic may be seen as an object-oriented extension of binary DATALOG with object methods, set expressions and negation. The following expression is the logic part of a CLAIRE rule:

```
edge(x,y) | exists(z, edge(x,z) & path(z,y))  =>  x.path :add y
```

The first extension is the introduction of interpreted functions, which are represented by object methods, with a model similar to [KW89]. These methods are either comparison methods or binary operations. For computing `x.tax` from `x.salary`, we may write:

```
exists(z, z = x.salary & z > 30000 & z < 50000 & y = 3000 + (z - 30000) * 0.28 )
    =>  x.tax := y
```

The second extension is the introduction of set expressions (*à la* LDL [NT89]). Here is a simple example that uses a set expression to represent the set of children who are still minors:

```
z = 2 + size({y in Person | y ∈ x.children & y.age <= 18})
    => x.family_size := z
```

The last extension is the introduction of negation (using `not(P(x,y))`). Although this object-oriented logic supports a top-down complete resolution strategy (with stratified negation), it is only used in CLAIRE to provide forward-chaining inference rules. A rule is defined by linking a logical condition to a conclusion, which is any CLAIRE expression. Here is a simple example:

```
compute_salary(x:Person[age:(18 .. 40)]) :: rule(
      x.salary < average_salary({y in Person | x.dept = y.dept})
    => x.next_salary := x.salary * 1.1)
```

To complete this brief overview of CLAIRE, we need to mention the "versioning" mechanism. A version (also called a *world*, using the AI terminology) is a virtual copy of the state of the objects. The goal is to be able to return to a previously stored state if a hypothetical branch fails during problem solving. As a consequence, worlds in CLAIRE are organized into a stack and only two operations are allowed: one for copying the current state of the database and another for returning to the previous state. The part of the objects that supports these defeasible updates is defined by the programmer and may include slots, tables, global variables or specific data structures such as lists or arrays (*defeasible* is an AI term that means "backtrack-able").

Each time we ask CLAIRE to create a new world, CLAIRE saves the status of defeasible slots (and tables, variables, ...). Worlds are represented with numbers, and various operations are provided to query the current world, create and return to previous worlds, with or without "forgetting" the most recent updates. The use of worlds is also encapsulated with the `branch(e)` control structure which creates a world, evaluates the expression `e` and returns `true` if the result of `e` is `true`, and otherwise backtracks to the previous world and returns `false`. Using these programming features makes writing search algorithms, such as branch-and-bound, much simpler in CLAIRE (or another language with search primitives such as ALMA [AS97]) than with traditional imperative languages such as Java or C++. For instance, here is a simple program fragment that solves the "n queens" problem:

```
D :: (1 .. SIZE)
column[n:D] : D := unknown
possible[x:D, y:D] : boolean := true
event(column)
store(column, possible)
r1(x:D, y:D) :: rule(exists(z:D, column[z] = y)   =>  possible[x,y] := false)
r2(x:D, y:D) :: rule(exists(z:D, column[z] + z = y + x)  =>  possible[x,y] := false)
r3(x:D, y:D) :: rule(exists(z:D, column[z] - z = y - x)  =>  possible[x,y] := false)
```

```
queens(n:(0 .. SIZE)) : boolean
    -> (if (n = 0) true
        else exists(p in D |
                    (possible[n,p] & branch( (column[n] := p, queens(n - 1)) )))
```

In this program, queens(n) returns true if it is possible to place n queens. The search tree (these embedded choices) is represented by the stack of the recursive calls to the method queens. At each level of the tree, each time a decision is made (an assignment to the column table), a new world is created so that we can backtrack whether this hypothesis (this branch of the tree) leads to a failure.

## 3. Set-Based Programming

### 3.1 Sets in CLAIRE

We gave a brief overview of the use of sets in CLAIRE in the previous section. However, there is more to the concept of "Set-Based Programming". For us, it comes from the combination of two principles:

- Sets are "first-class citizens" with a choice of multiple representations (abstraction);
- Abstract sets may be used to capture "design patterns" (in the sense of the Object-Oriented Programming community, that is, a motif that repeats itself in a program because it represents an abstract operation).

The first aspect in CLAIRE is illustrated by the variety of data structures that are available to represent collections of objects. These include classes, arrays, lists, sorted collections or bit-vectors. Also, we already mentioned that any type can be used as a collection, as illustrated in the following example:

```
for x in person[age:(15 .. 18)]  print(x)
```

Moreover, new collection classes can be introduced. In the following example, we introduce collections that are implemented with a hash table (hashing an object to get an index in a list is a standard method).

```
Hset<of> <: collection(of:type, content:list, index:integer)
add(s:Hset<X>, y:X) : void  ->  let i := hash(s.content, y) in ( ... )
set!(s:Hset) : set  ->  {x in s.content | known?(x)}
```

In addition to concrete sets, CLAIRE supports set expressions, which can be used freely since they are built lazily by the compiler so that any useless allocation and construction is avoided, as we shall see later. We already mentioned set construction by image or selection. New set expressions can also be introduced through the concept of pattern. A pattern is a function call that can be lazily dealt with. The first step is to define a new set operator, such as:

```
but(x:collection,y:any) : set  ->  {z in x | z != y}          // (x but y) ⇔ x \ {y}
```

Now we can treat the pattern (s but x) lazily without building a concrete data structure that would represent it. The pattern type F[tuple(X,Y)] is the set of function calls with F as the selector (function) and two arguments (stored in the args list) of respective types X and Y. For instance, we can declare how to compute membership on the but pattern so that the compilation of the expression (x ∈ (Person but john)) will be performed by code substitution into (x ∈ Person & x != john).

CLAIRE supports true abstraction for collections since one may substitute a representation by another and not have to change a line of code in generic methods, such as the following example:

```
sum(s:subtype[integer]) : integer            // sum is an inline method (=> is the syntactic marker for inline)
    => let d := 0 in (for x in s  d :+ x,  d)
```

On the other hand, re-compiling will be necessary since the efficient implementation of abstraction relies on source-to-source transformation.

### 3.2 Customizable Iteration

One of the most important aspects of set programming from a practical perspective is the ability to customize set iterations. Iteration plays an important role in CLAIRE, either in an explicit form (using the for control structure) or in an implicit form (an iteration is implied in the image/selection set expressions as well as in the some/exists structures) as shown in these examples:

```
{x in (1 .. 10) | f(x) > 0}
{size(c) | c in (class but class)}         // note: class is a class
```

For each of these iterations, the compiler uses source-to-source optimization and generates the equivalent, but faster, expressions. The first one is compiled into:

```
let s := {}, x := 1 in  (while (x <= 10)  (if (f(x) > 0)  s :add x,  x :+ 1), s)
```

Since the set of collection classes is extensible, the compiler supports the extension of the iteration mechanism through the definition of the `iterate` method. For any expression *e(v)* that contains a variable v, *iterate(x,v,e)* is an expression that produces *e(v)* to be evaluated for all values v in the set *x*. Here what we would write if *x* is a Hset:

```
iterate(x:Hset, v:Variable, e:any) : any  =>  for v in x.content  (if known?(v)  e)
```

This is also true for patterns, thus we can explain how the iteration of a pattern should be compiled lazily without building the set:

```
iterate(x:but[tuple(collection,any)],v:Variable,e:any) : any
       => for v in args(x)[1]  (if (v != args(x)[2])  e)
```

This mechanism has also been used to implement the lazy iteration of image or selection sets. As a consequence, an expression such as:

```
for x in {f(x,0) | x in (myHset but 12)}  print(x)
```

will be transformed into:

```
for x in myHset.content  (if (known?(x) & x != 12)  print(f(x,0))
```

As a consequence, it is both natural and efficient in CLAIRE to use "virtual" sets that are iterated as a programming pattern. This yields programs that are easier to read since we separate the filtering of the data from the processing. Note that the *iterate* optimizing pattern does not carry a type since the type of an iteration will be determined dynamically after the substitution as occurred.

### 3.3 Sets and Design Patterns

This section illustrates the previous feature through two examples of using sets as a design and programming pattern. The first example is the concept of embedded (doubly) linked lists. Linked lists are very common and useful objects, since they provide insertion and deletion in constant time. The implementation provided by most libraries is based on cells with pointers to next/previous cells in the list. This is useful but has the inconvenience of requiring dynamic memory allocation. When performance is critical, most programmer use embedded linked list, that are implemented by adding the two `next`/`previous` slots to the objects that must be chained. This only works if few lists (e.g., one) need to be kept, since we need a pair of slots for each kind of list, but it has the double advantage of fewer memory access (twice fewer) and no dynamic allocation as the list grows. For instance, if we have objects representing tasks and if we need a list of tasks that have to be scheduled, we may define the class `Task` as follows:

```
Task <: object( ..., next:Task, prev:Task, ...)
```

The obvious drawback is the loss of readability, since the concept of linked list is diluted into pointer chasing. In CLAIRE, we can use a pattern to represent the concept of a linked chain using these two slots as follows. We first define `chain(x)` which is the list of tasks obtained by following the `next` pointer:

```
chain(x:Task) : list<Task>
      -> let l := list<Task>(x) in                    // x is the first member
           (while known?(next, x)  (x := x.next, l :add x),  l)
```

Note that the list constructor carries the type of the new list when we used strongly typed lists. `list<integer>(1,2,3)` is the list with 3 element (1,2 and 3) whose member type is integer.

We define the iteration of the `chain` pattern in a very similar way:

```
iterate(x:chain[tuple(Task)], v:Variable, e:any) : any
       => let v := args(x)[1] in
            (while true (e,                           // (a,..,c) is a sequence of exprs
                         if known?(next, v)  v := v.next  else break()))
```

We can now use the "abstract" chains as any other collection in CLAIRE:

```
sum({weight(x) | x in chain(t0)})
```
```
count(chain(t1))
```

The interest of this approach is to keep the benefit of abstraction at no cost (we do not see the pointer slots anywhere and we can later replace embedded linked list with another representation). The code produced by the optimizer is precisely the pointer-chasing loop that we would have written in the first place.

The second example shows the power of combining patterns with the concept of iterators. Iterators are a very popular solution for iterating collection in most Java and C++ libraries. However, for most collection types they induce a needless overhead (thus, we do not use them in CLAIRE). However, there are more complex data structures for which iteration can be seen as traversing the structure, and for which the iterator embodies the traversal strategy. An iterator is simply defined by two methods, `start` and `next`, which respectively provide the entry point for the iteration and the succession relationship. The following is an example of a tree iterator. We introduce a pattern (T by I) which produces the set of nodes in T using the traversing strategy of the iterator I:

```
Tree <: object(value:any, right:Tree, left:Tree)

TreeIterator <: object(tosee:list, status:boolean)

iterate(x:by[tuple(Tree,TreeIterator)], v:Variable, e:any) : any
    => let v := start(args(x)[2], args(x)[1]) in
         while (v != unknown)  (e, v := next(args(x)[2], args(x)[1]))
```

We can define many types of iterators corresponding to the various popular strategies for iterating a tree (Depth-First Search, Breadth-First Search, etc.):

```
TreeIteratorDFS <: TreeIterator()              // Depth-First Search strategy
DFS :: TreeIteratorDFS()
start(x:TreeIteratorDFS, y:Tree) -> ...
next(x:TreeIteratorDFS, y:Tree) -> ...
```

We may then use the pattern (T by I) as any other collection and simply write:

```
sum({y.weight | y in (myTree by DFS)})
```

which associated optimized code is the following:

```
let d := 0, y := start(DFS, myTree) in
    (while (y != unknown)  (d :+ y.weight, v := next(DFS, myTree)), d)
```

The iterator object plays two roles: its type will determine which sort of tree traversal is required and it also contains the state of the traversal (the *tosee* and *status* slots) that is exploited by the *start* and *next* method. Thus, if the same iterator may be used on different trees, the simultaneous iteration of many trees will require many iterators.

## 4. Rule-Based Programming
### 4.1 Claire Rules
#### 4.1.1 A Logical Language for Assertions

A rule in CLAIRE is obtained by associating a logical condition to an expression. Each time the condition becomes true for some objects, the expression will be evaluated for these objects (represented by the rule's free variables). A condition is defined with a logical language that includes *predicate logic* to describe the state of one or many objects and *event logic* to capture the occurrence of a precise event.

To define a logical condition, we use the logic language defined by the following grammar.

```
<assertion> ≡ <expression> <comp> <expression>  |  <variable> :: <class>  |
              <value> := < <expression> | (<expression> <- <variable>) >  |
              exists(<variable>, <assertion>)  |  not(<assertion>)  |
              if (<expression> <comp> <expression>) <assertion> else <assertion>  |
              <assertion> & <assertion>  |  <assertion> | <assertion>

<expression> ≡ <variable>  |  <entity>  |  <function>(<expression>)  |
               <expression> <operation> <expression>  |
               {<var> in <expression> | <assertion>}  |
               list{<var> in <expression> | <assertion>}

<value>     ≡  <table>[<expression>]  |  <expression>.<property>

<rule>      ≡  <name>(⟨<variable>:<type>⟩⁺) :: rule(<assertion> => <statement>)
```

The value of a logical expression is a CLAIRE entity. Logical assertions are made by combining expressions. The most common type of assertion is obtained by comparing two expressions with a comparison operation. A comparison operation is an operation that returns a Boolean value.

The novelty in CLAIRE 3.0 is the introduction of event logic. There are two events that can be matched precisely: the update of a slot or a table, and the instantiation of a class. The expression x.r := y is an event expression that says both that x.r = y and that the last event is actually the update of x.r from a previous value. Thus, the introduction of event expressions serves two purposes:

- control the triggering of a rule more precisely by telling which event should be considered. Consider the following example:

    ```
    x.age = y  &  z.age = 2 * y
    ```

    This logical expression may become true (and trigger an associated rule) from an update of the age slot in two manners, either because x.age has changed or because z.age has changed. In most cases, this behavior is precisely what is expected. In contrast, the following expression :

    ```
    x.age := y & z.age = 2 * y
    ```

    adds to the previous expression the explicit constraint that the triggering update is applied to the object x.

- use the "previous value" of the updated expression x.r in the logic expression. This is achieved with the expression x.r := (y <- z) which says that the last event is the update of x.r from z to y. For instance, here is the event expression that states that x.salary crossed the 100000 limit:

    ```
    x.salary := (z <- y) & y < 100000 & z >= 100000
    ```

The other event that can be described with the event logic is the instantiation of a class. The expression x :: C means that x is a new instance of the class C. As previously, the event expression is the combination of the statement x ∈ C and the additional constraint that the last event is precisely the creation of the new instance x of the class C. We use object instantiation to represent external events when writing an application that must be integrated with other components using a *publish & subscribe* event model, such as the Java Bean model [Eng97].

The value of a logical assertion is always a Boolean, thus logical assertions can be combined with & (and) and | (or). In addition to pre-defined operations such as <= or +, it is possible to use new properties inside logical assertions but they need to be described using the *description* table and a set of keywords including *comparison, binary_operation, monoid, group_operation*. By declaring the algebraic characterization of a property we are telling the rewriting system that produces the algebraic representation how to handle the logic expression.

To define a rule, we define a list of free – universally quantified – variables, that are introduced as parameters of the rule, a condition, which is given as an assertion using the previously defined variables and a conclusion that is preceded by =>. Here is a classical transitive closure example:

```
r1(x:Person, y:Person) :: rule(
      exists(z, x ∈ z.friends & z ∈ y.friends)
      => y.friends :add x )
```

### 4.1.2 Event-Based Rules

Rules are checked (efficiently as we shall later see) each time that an *event* occurs. Events come in two kinds. The first kinds of events are updates to slots or tables that are declared as event generating with the *event* statement.

```
<events definition>  ≡  < event | noevent >( < <table> | <property> >* )
```

The second kinds are instantiations of classes. These include the instantiation of "event objects" classes that represent "external events" used in a larger-scale application, but also the instantiation of local classes. Note that the "instantiation event" occurs after the initialization of the object (with default or given values for the slots).

A rule considers as events all slots or tables that have been declared as such *before* the rule has been declared, so the event declaration needs to precede the rule declaration, in order to create the "demons" that will watch over the given relation and fire the rule when needed. The declaration noevent(...) may be used to prevent explicitly a rule to react to some relations that were declared previously as events for other rules. By spreading the event declarations (and occasionally noevent declarations) before and between the rules, one has a precise control over the triggering of the rules. This level of control is further refined with the use of event expressions within the condition of the rule.

Updates that are considered as events are:
- x.r := y, where r is a slot of x and event(r) has been declared.
- a[x] := y, where a is a table and event(a) has been declared.
- x.r :add y, where r is a multi-valued slot of x (with range bag) and event(r) has been declared.
- a[x] :add y, where a is a multi-valued table and event(a) has been declared.
- The instantiation of a class C, whether static (x :: C(…)) or dynamic (new(C,…)).

The handling of mono- vs. multi- valuation is an important aspect of this logic. Multi-valued slots in CLAIRE are slots that contains multiple values (i.e., friends is multi-valued since I may have many friends). A multi-valued slot is implemented using a container (a list or a set), however, from a semantic point of view, we are only interested with the multi-valued binary relationship between a person and her friends. Operations on a multi-valued slots are the addition or the deletion of a new member, whereas a mono-valued slots only supports the update the value (replacing a value by another one). Thus, CLAIRE makes a difference between a multi-valued slot (for instance, a slot `friends` with range `set<person>`) and a mono-valued slot with range `set` (we can use the same example). *The management of inverses is different*: in the first case, the inverse of `friends` is a relation from `Person` to `Person`; in the second case, it is a relation from `set<Person>` to `Person`. Similarly, the atomic events are different: in the first case it is the addition of a new `friend`; in the second case, it is the replacement of a set of `friend` by another set. It is our experience that both modeling options are needed to capture a large set of situations.

### 4.1.3 Examples

The following first example is taken from our cumulative scheduling program presented in [CL96b]. Cumulative scheduling is a special planning problem where some tasks have to be scheduled on resources. A resource may handle several tasks at the same time, but is limited by a finite capacity (`capacity`). The problem is to perform all tasks such that the total completion time is minimal. The amount by which a task "consumes" a resource must fit between two bounds (`minUse/maxUse`).

The rule `eject-up` (see below) involves two data structures: `Task` and `Interval`. For each `Task` t, we store (in slots) the required resource `t.res`, time bounds `t.earliestStart` and `t.latestEnd`, the associated work `t.work`, and its extremal resource use levels `minUse/maxUse`. For each `Interval` S, we store its bounds (`Left/Right`), the common resource of all tasks in the interval (`Res`), and the sum of the work amount of tasks (`Work`) in the `Interval`.

The rule considers a task t and an interval S. Let A be the earliest start time of the lower bound of S, B the latest end time of the upper bound of S, and a the earliest start time of t. Whenever t and S are such that:

- they share some common `resource` r that is actually cumulative (i.e. r.capacity > 1),
- S is not an empty `Interval`, but t is not included in S,
- t cannot start before A and end before B ((t.work − max(0,A − a)) * t.maxUse > slack), because there is not enough slack, even with the maximal consumption level for t, in order to perform as much as possible of t before A.
- t cannot start before A and end after B ((B − A) * t.minUse > slack) because there is not enough slack, even with the minimal consumption level for t.

In such a case, the rule deduces that t necessarily starts after A, and analyses the amount of t that can be performed before B in order to update the time bounds of t. We use two update methods `increaseEarliestStart` and `increaseLatestStart` in order to maintain the `Interval` structure.

```
[eject-up(t:task, S:Interval) ::
    exists(r:resource, S.Res = r & t.res = r & r.capacity > 1 &
        not(empty(S)) & not(contains(S,t)) &
        exists(a, a = t.earliestStart &
           exists(A, A = S.Left.earliestStart &
              exists(B, B = S.Right.latestEnd &
                 exists(slack, slack = (B - A) * r.capacity – S.Work &
                              (t.work – max(0,A - a)) * t.maxUse > slack &
                              (B - A) * t.minUse > slack )))))
    =>  increaseEarliestStart(t, B - (slack div+ t.minUse)),
        increaseLatestEnd(t, B - (slack div+ t.maxUse) + t.minDuration) ]
```

This rule is triggered whenever a time bound `earliestStart`/`latestEnd` or the `contains` relation is modified. There are multiple propagation patterns and writing the equivalent procedural code is really a tedious and error-prone task.

We will now present a few examples that rely on the event-based logic that we have introduced. First, here are three (simplified) examples taken from the field of constraint propagation. The first one tells that if two tasks are consecutive (attached means without idle time in between), we can propagate the increase in `minStart` (earliest starting time) directly. The second tells how to propagate a `duration` increase.

```
R1(x:task, y:task, z:integer, u:integer) :: rule(
    x.attached = y & x.minstart := (z <- u)   =>   y.minstart :+ (z – u) )

R2(x:task, I:taskInterval, z:integer, u:integer) :: rule(
    x.minStart >= I.minstart & x.maxEnd <= I.maxEnd & x.duration := (z <- u)
    => I.duration :+ (z – u) )

R3(x:task, I:taskInterval, z:integer, u:integer) :: rule(
    x.minStart := (z <- u) & z >= I.minstart & u < I.minStart & x.maxEnd <= I.maxEnd
    => I.duration :+ x.duration )
```

The first two rules are similar and illustrate the use of the event pattern to capture a transition and propagate an incremental change. The second rule also shows that the event pattern forces the rule to trigger when a change is made on the duration attribute, which is what we intend here, since a different rule (R3) handles the case when a change is made on the minStart attribute. In R3, the event pattern is used to capture the state transition (x now belongs to the task interval and did not before).

The next two examples illustrate the use of event-rules to model state-transition systems. They are taken from an "artificial life" scenario where units move from cell to cell. The first rule says that if a unit moves to a position where another unit is already installed, a conflict occurs which will result into a failure of the incoming unit if its strength is less than the strength of the residing unit. The second re-computes the visibility graph for a unit that gets "out-of-the-woods".

```
RS1(x:unit, y:unit) :: rule(
    x.location := y.location & x.strength <= y.strength  =>  failure(x))
RS2(x:unit, z:position, u:position) :: rule(
    x.location := (z <- u) & u.forest? = true & z.forest? = false
    => computeVisibility(x) )
```

The use of event logic enables us to represent the dissymmetry between x and y in the first rule, which would not be possible with sole predicate logic. The second rule is another example of using the event pattern to capture a state transition (from being in-the-woods to out-of-the-woods). The use of "?" after forest has no special meaning in CLAIRE, it is a common practice to add "?" to names of slots or methods that represent a Boolean.

## 4.2 Algebraic Representation of Rules

### 4.2.1 Principles of Algebraic Rules

CLAIRE rules are compiled into procedural demons in three phases (cf. Figure 1). First, the condition part of the rule is rewritten into an algebraic expression, which is next differentiated into many functions that are finally transformed into *if-write* demons. The process of rule compilation relies on the use of a relational algebra. This algebra $A(R)$ contains relational terms (elements of R) that represent binary relations; it is generated from a set of constants (Cartesian products of types), of variables (CLAIRE slots and tables) and a set of operators (similar to [McL81]). These operators can define unions, intersections and compositions over relational terms, but also more advanced algebraic functions such as the composition of a binary operation or a comparison operation with two binary relations, as we shall see in the next section.

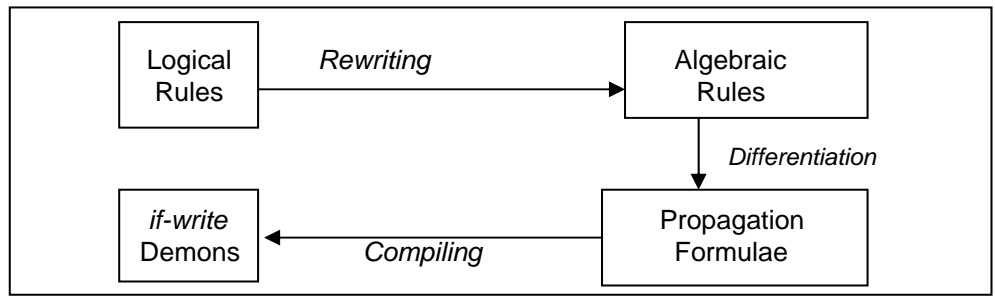

*Figure 1. Compilation scheme of CLAIRE rules.*

We will briefly describe each step of the compilation process, and follow the transformation of a rule, from its logical expression towards the generated procedural demons. Here, we use the closure rule, which performs a transitive closure over a graph, described through the binary relations edge and path:

```
closure(x:point, y:point) :: rule(
    edge(x,y) | exists(z, edge(x,z) & path(z,y))  =>  x.path :add y )
```

- The first phase of the compilation process is the transformation of the conditional part of the rule into a relational formula that belongs to $A(R)$. Translation into the algebraic form is based on rewriting and involves a lot of knowledge about object methods. The principle is to solve the equation assertion(x,y), while considering that x is known and that y is sought. The result is the relational algorithm that explains how to get y from x, which is represented as a term in $A(R)$. The conditional part or the rule closure is transformed into the following algebraic term t1:

    t1 = edge ∪ (path o edge)

where ∪ and o are respectively the union operator and the composition.

- Next, a phase of symbolic differentiation [SZ88] computes the effect of adding a pair of values into a table or a slot. The differentiation is a function that is applied to the algebraic term `t`, with respect to each relation R occurring in the condition, and that produces a term $\partial t/\partial R$ in $A(R \cup \{\mathbb{0},\mathbb{1}\})$. $\mathbb{0}$ and $\mathbb{1}$ are two constant functions that represent respectively the empty function ($\mathbb{0} : (x,y) \to \varnothing$) and the identity function ($\mathbb{1} : (x,y) \to \{(x,y)\}$). A set of differentiation rules (one for each operation on binary relations) is used to compute the term $\partial t/\partial R$. The term `t1` computed above is differentiated with respect to the relation `edge` and yields the following function `f1`:

    `f1 = ∂t1/∂edge = 𝟙 ∪ (path o 𝟙)`

    or in other terms:

    `f1 : (a,b) → {(a,b)} ∪ {(a,y) | (b,y) ∈ [path]}`

- The last phase of the compilation of a rule consists of associating the set of functions (derived from each relational term occurring in the condition) and the CLAIRE expression in the conclusion, so as to build one *if-write* demon per function. The demon is used to propagate each update so that rules are triggered in an incremental manner, yielding an approach similar to [SKG87].

We now examine these different phases with more details in the following sections.

### 4.2.2 A Relational Algebra

The relational algebra contains terms that represent binary relations. It is generated from a set of constants (cartesian products of types), a set of variables (the relations that are described with rules) and a set of relational operators. This set may be described as follows :

- $\cup, \cap$, **o** are the usual operators for union, intersection and composition (binary join).
- $\psi[+](r_1, r_2)$ represents the composition of a binary operation ($+$) with two binary relations $r_1$ and $r_2$.
- $\phi[<](r_1,r_2)$ is the composition of a comparison operation ($<$) with two binary relations $r_1$ and $r_2$.
- $if[\phi[<](r_1,r_2)](r_3,r_4)$ is the conditional composition of a test ( $\phi[<](r_1,r_2)$ ) with two binary relations $r_3$ and $r_4$ which represent the two branches of the conditional ($r_3$ if the test succeeds and $r_4$ otherwise).
- $\{z{:}r_1, r_2\}$ represents a relation that associates to an object the set of objects $z$ bound through $r_1$ which also "satisfy" $r_2$ (hence, $r_2$ is a filtering relation - a $\phi$ or a composition of $\phi$'s). This operator is introduced to handle set creation in logical formulae.
- Set_expansion(r), written $r^*$, is the reciprocal operator which is used to treat a relation r with range *set* as a multi-valued relation.

Here is the formal definition of the relational operators:

$$(x,y) \in r_1 \cup r_2 \Leftrightarrow (x,y) \in r_1 \vee (x,y) \in r_2$$
$$(x,y) \in r_1 \cap r_2 \Leftrightarrow (x,y) \in r_1 \wedge (x,y) \in r_2$$

$$(x,y) \in r_1 \mathbf{o}\, r_2 \Leftrightarrow \exists z, (x,z) \in r_2 \wedge (z,y) \in r_1$$
$$(x,y) \in \psi[+](r_1, r_2) \Leftrightarrow \exists z_1, z_2, (x,z_1) \in r_1, (x,z_2) \in r_2 \wedge y = z_1 + z_2$$
$$(x,y) \in \phi[<](r_1, r_2) \Leftrightarrow x = y \wedge \exists z_1, z_2, (x,z_1) \in r_1, (x,z_2) \in r_2 \wedge z_1 < z_2$$
$$(x,y) \in if[\phi[<](r_1,r_2)](r_3,r_4) \Leftrightarrow$$
$$\{(x,y) \in r_3 \wedge \exists z_1, z_2, (x,z_1) \in r_1, (x,z_2) \in r_2 \wedge z_1 < z_2 \} \vee$$
$$\{(x,y) \in r_4 \wedge \exists z_1, z_2, (x,z_1) \in r_1, (x,z_2) \in r_2 \wedge \neg(z_1 < z_2) \}$$
$$(x,y) \in \{z{:}r_1, r_2\} \Leftrightarrow y = \{z \mid (x,z) \in r_1 \wedge (z,z) \in r_2\}$$
$$(x,y) \in r^* \Leftrightarrow \exists z, (x,z) \in r \wedge y \in z$$

To achieve the same expressive power as the logical fragment that we use to define conditions on objects, we need to introduce the ability to share a sub-term, which will correspond to the introduction of existential variables. Let us now define $A(R)$ the term algebra generated by the previous set of operators, the constant cartesian products of types (such as *person × {0,1,2}*) and the set R of relational variables (that represent relations from CLAIRE, which are tables or properties). We extend our definition as follows:

- If x is a relation variable that does not belong to R, if $t_2(R_1,...,R_n,x)$ is a term from $A(R \cup \{x\})$ and $t_1(R_1,..,R_m)$ is a term from $A(R)$, then $\lambda(x{:}t_1).t_2$ is a term from $A(R)$ which represent the relation denoted by $t_2$ in an new "object system" where x is bound to the relation denoted by $t_1$.

Notice that an "object system" is simply defined by the valuation function $q$ that associates to each relation name in R the actual binary relation in the current system. We write $[t]_q$ the binary relation denoted by the term $t$ for a given valuation $q$. A key result in [Cas94] is that the logic fragment that we use and the relational algebra $A(R)$ have the same expressive power. This means that for any condition $C$ with at most two free variables $x$ and $y$ there exists a term $t$ of the algebra such that $C(x,y) \Leftrightarrow (x,y) \in [t]_q$. The term is not necessarily unique and the difficulty is to produce an "optimized" term, which represents the straightest relational computation that is necessary to produce the set of $y$ starting from the value of $x$.

The transformation of a logic rule into its algebraic representation is based on rewriting, as explained in [Cas94]. This is achieved with rewriting rules that incorporate the algebraic knowledge about objects' methods such as:

$$group(+,0,-) \ \& \ x + y = z \Rightarrow x = z - y$$
$$monoid(*,|,/) \ \& \ x * y = z \Rightarrow (y | z) \ \& \ (x = z / y)$$

The first rule states that if we have a Group (+,0,-) the equation x + y = z can be rewritten into x = z – y where – is the inverse operation of + (thus, this applies to matrices as well as integers). The second rule states how to rewrite the same equation for a weaker structure: here * has no complete inverse operation but a partial one, /, associated by an inversibility relation – divisibility.

In the previous years, we have added the concept of simplification rules, which implement optimizing strategies (replacing an algebraic term by another one with the same denotation but which is more efficient to compute) in a manner similar to the optimization of SQL queries. The rules are applied recursively while algebraic terms are being built. They are mostly based on three ideas:

(1) filtering terms should be ordered from simpler to more complex, and applied as soon as possible from an evaluation order point of view,

(2) constants sub-terms must be recognized, simplified, and re-organized. For instance, linear sub-term factorization (2 * x + 3 * x $\Rightarrow$ 5 * x) is applied through this simplification step.

(3) inversible sub-terms must be detected because they can be used (in their inverse form) to avoid costly joins.

The motivation for this translation is that we are able to represent propagation as a formal computation over algebraic terms, called *differentiation*. Suppose that $t$ is a term for $A(R)$, $R_i$ a relation variable from R and $q$ a valuation function representing the current object system. We now consider an atomic update, that is the addition of a pair $(x,y)$ to the relation $q(R_i)$. The term $t$ now represents a new relation in the new object system $q'$. The goal of differentiation is to compute $[t]_{q'} - [t]_q$, that is the set of object pairs that "appeared" in the relation represented by $t$ "because" of the update (since we use a monotonic algebra, we have $[t]_{q'} \supset [t]_q$).

Let us define the functional algebra $F(R)$ as $A(R \cup \{0, 1\})$, where each relation of R is mapped to a constant function. We use two constants $0, 1$ that represent respectively the "empty function" $((x,y) \to \emptyset)$ and the unit function $((x,y) \to \{(x,y)\})$. To represent the propagation of an update in a set-valued expression, we also need to extend $F(R)$ with a new projection operation $\pi(t)$ defined as follows.

$$(x,y) \in [\pi(t)(a,b)]_q \Leftrightarrow x = y \wedge \exists z, (x,z) \in [t(a,b)]_q$$

Thus, $\pi(t)$ is a the first projection of the relation represented by $t$.

Differentiation is a formal operation on $A(R)$ which associates to a pair $(t,R_i) \in A(R) \times R$ a term $\partial t/\partial R_i$ from another functional algebra $F(R)$ (which represents a function that associates a binary relation to an object pair) such that :

$$[t]_{q'} - [t]_q \subseteq [\partial t/\partial R_i(a,b)]_{q'} \subseteq [t]_{q'} \tag{1}$$

To guarantee the interest of differentiation we also impose that $\partial t/\partial R_i(a,b)$ is the smallest term from $F(R)$ which satisfies (1) according to the partial order induced by inclusion. However, (1) is a necessary and sufficient condition to prove the validity of using differentiation to model propagation.

### 4.2.3 Extending the Algebra to Handle Event-based Logic

The algebra is extended with two operators:
- $\chi(R)$ is an "event-filter" : it represents the same relation as R, but it will behave differently when the term is differentiated.
- R' represents the "previous" version of the relation R before an atomic update. It cannot be used freely in the relational algebra, which would require a much more complex implementation, but only in the following pattern:
    $$\lambda(z:R').t\{\chi(R)\}$$
  where t{$\chi$(R)} is a term which contains $\chi$(R) as a sub-term.

The differentiation rules are straightforward:

$$\partial\chi(R)/\partial R = 1 \ , \ \partial\chi(R')/\partial R = 0$$
$$\partial\chi(\lambda(z:R').t\{\chi(R)\})/\partial R = \lambda(z:R'). \ \partial t/\partial R$$

The compilation rules (cf. [Cas94]) are two methods, *expR* and *expF*, that represent the semantic of the operators through code generation. Both are defined by structural induction for each operator.

- `expR`(t,x,y,e(y) )   is a code fragment that applies the expression e(y) to each object y that is linked to x according to the binary relation represented by the term t. Here is a simple example for a binary composition:
    $expR(t_1 \ \mathbf{o} \ t_2, x, y, e(y)) = expR(t2, x, z, expR(t_1,z,y,e(y)))$

- `expF(∂t/∂R,x,y,u,v,e(u,v) )` is a code fragment that applies the expression e(u,v) for each pair (u,v) that belongs to the relation which is the value of [∂t/∂R](a,b), where we recall that ∂t/∂R is a term of a functional algebra that represent a function that associates a binary relation to each pair of objects. Here is another example for compiling the composition of a derivative and a relation:

$$expF(∂t/∂R \; \mathbf{o} \; t, x,y,u,v,e(u,v)) = expR(∂t/∂R, x,y,a,v, expR((t)^{-1}, a,u,e(u,v)))$$

This last example shows that although differentiation rules are simple and intuitive, their actual implementation through code generation may be more subtle. The ability to inverse any term from the algebra is not easy to reach, although many terms may be inversed through recursive formulas. For the few but annoying exceptions, we actually revert to the logical expression and re-apply a translation step (rewriting) while changing the variable ordering (each translation step takes one free variable as the "base" and the other one as the target). To achieve this, we need to ensure that for each composition sub-term ($r_1$ **o** $r_2$), the second relation $r_2$ is always applied to this "base" variable. We use the associativity of binary join to generate terms like ($r_1$ **o** ($r_2$ **o** $r_3$)) as opposed to (($r_1$ **o** $r_2$) **o** $r_3$).

Similarly, the rule for differentiating a "lambda-term" (($\lambda z.R$).t) (i.e., the binding of a sub-term t to a variable z) looks simple:

$$\partial(\lambda(z:t_1).t_2)/\partial R \;=\; \lambda(z:\partial t_1/\partial R).t_2 \;\cup\; \lambda(z:t_1).\partial t_2/\partial R$$

However, the actual evaluation of the second part $\lambda(z:t_1).\partial t_2/\partial R$ is subtle, since we need to know the value of this base variable before being able to apply the sub-term $t_1$ to find *z* (and compute the whole expression). This requires a recursive traversal of the derived term to find out exactly when the base variable can be computed.

These compilation methods, `expR` and `expF`, are extended to the new operators as follows:

$$expR(\chi(R), x, y, e(y)) = \text{nil}$$
$$expF(\lambda(z:[R]).(\partial t/\partial R), x,y,u,v,e(u,v))) = \text{let } z = \text{LAST in } expF(\partial t/\partial R, x,y,u,v,e(u,v))$$

The first formula tells that $\chi(R)$ will be ignored in a rule if it is not differentiated precisely according to R. Thus it acts as a filter. The second formula shows that we make a direct reference to the previous value LAST of the attribute that is being updated. This value is stored when we perform the update as we shall later see.

The following is the set of derivation rules for this new logic.

---

$\forall (i,j), \; \partial R_j/\partial R_i = \mathbb{1}$ if i = j and $\mathbb{0}$ otherwise.

$\forall (i,j), \; \partial R_j^{-1}/\partial R_i = \mathbb{1}^{-1}$ if i = j and $\mathbb{0}$ otherwise.

$\forall p \in P, \; \partial p/\partial R_i = \mathbb{0}$.

If $S_1$ and $S_2$ are two types, $\partial(S_1 \times S_2)/\partial R = \mathbb{0}$

$\forall t_1, t_2 \in A(R), \; \partial(t_1 \; \mathbf{o} \; t_2)/\partial R_i = (\partial t_1/\partial R_i \; \mathbf{o} \; t_2) \cup (t_1 \; \mathbf{o} \; \partial t_2/\partial R_i)$

$\forall t_1, t_2 \in A(R), \partial(t_1 \cap t_2)/\partial R_i = (\partial t_1/\partial R_i \cap t_2) \cup (t_1 \cap \partial t_2/\partial R_i)$

$\forall t_1, t_2 \in A(R), \partial(t_1 \cup t_2)/\partial R_i = (\partial t_1/\partial R_i \cup \partial t_2/\partial R_i)$

$\forall t_1, t_2 \in A(R), \forall f \in F_O, \; \partial \psi[f](t_1, t_2)/\partial R_i = \psi[f](\partial t_1/\partial R_i, t_2) \cup \psi[f](t_1, \partial t_2/\partial R_i)$

$\forall t_1, t_2 \in A(R), \forall \theta \in F_C, \partial \phi[\theta](t_1, t_2)/\partial R_i = \phi[\theta](\partial t_1/\partial R_i, t_2) \cup \phi[\theta](t_1, \partial t_2/\partial R_i)$

$\forall t_1, t_2 \in A(R), \forall z \in V, \partial(\lambda(z:t_1).t_2)/\partial R_i = \lambda(z:\partial t_1/\partial R_i).t_2 \cup \lambda(z:t_1).\partial t_2/\partial R_i$

$\forall t_1, t_2, t_\phi \in A(R) \mid t_\phi = \phi[\theta](\ldots)),$

$\quad\quad \partial \text{if}[t_\phi](t_1, t_2) = \text{if}[\partial t_\phi/\partial R_i](t_1, t_2) \cup (\partial t_1/\partial R_i \; \mathbf{o} \; t_\phi) \cup (\partial t_2/\partial R_i \; \mathbf{o} \; \neg t_\phi)$

$\forall t \in A(R), \; \partial\{z:t_1, t_2\}/\partial R_i = \{z:t_1, t_2\} \; \mathbf{o} \; \pi \; (\partial t_1/\partial R_i \cup \partial t_2/\partial R_i)$

$\forall j, \; \partial (R_j^*)/\partial R_i = (R_j^*) \; \mathbf{o} \; \pi \; (\partial R_j/\partial R_i)$

$\forall r \in R, \; \partial \neg r/\partial r = \mathbb{0}$.

$\forall (i,j), \; \partial \chi(R_j)/\partial R_i = \partial R_j/\partial R_i$

$\forall t \in A(R), \; \partial(\lambda(z.[R]) \; t)/\partial R \;=\; \lambda(z.[R]) \; (\partial t/\partial R)$

---

Figure 2: *Differentiation rules*.

These rules completely specify differentiation as a formal computation. The analogy with rules for numerical differential calculus is no coincidence and can be proven using a matrix model for binary relations [Cas94].

### 4.2.4 Examples

For each rule given in Section 4.1.3, we give the equivalent algebraic term t and its derivation ∂t/∂R according to the appropriate relation. Since these rules use an event pattern to capture one precise pattern, there is only one relation R such that the derivative ∂t/∂R is not empty.

```
R1 [t1]:  λ(u:minstart').λ(z:χ(minstart)).attached
R2 [t2]:  λ(I:task x taskInterval).λ(z:duration).(χ(duration) o
                    φ[>=](minStart, minStart o I) o φ[<=](maxEnd, maxEnd o I))
R3 [t3]:  λ(I:task x taskInterval).(
                    φ[<=](maxEnd, maxEnd o I) o λ(u:minstart').
                    (φ[<](u, minStart o I)) o φ[>=](χ(minstart), minStart o I)))
RS1 [t4]: λ(y:(location⁻¹ o χ(location))). y o φ[>=](strength o y, strength)
RS2 [t5]: λ(u:location').(φ[=](forest? o u, false) o φ[=](forest? o χ(location), true)
```

It is easy, although tedious, to verify that these algebraic terms represent indeed the conditions that define rules. For instance, λ(y:(location⁻¹ o χ(location))). φ[>=](strength o y, strength) (x,z) ⇔ ∃ w, location(y) = w ∧ location(x) := w ∧ strength(y) >= strength(x). We then apply differentiation with respect to all relations that are involved in the rules/

Here are some of the derived terms:

```
∂t1/∂minStart = λ(u:minstart').((λz.𝟙 ) attached⁻¹ )
∂t2/∂minStart = λ(I:task x taskInterval).λ(z:duration'). (
                    𝟙 o φ[>=](minStart, minStart o I) o φ[<=](maxEnd, maxEnd o I))
∂t4/∂location = λ(y:(location⁻¹ o 𝟙)). φ[>=](strength o y, strength)
∂t5/∂location = λ(u:location').(φ[=](forest? o u, false) o φ[=](forest? o 𝟙, true)
```

Each derived term represents a function, which meaning is obtained from the semantic of the functional algebra. For instance, the first term is a function that takes a pair of objects (*a,b*), binds *z* to the value *b* (from the update *a.minstart := b*), binds *u* to the previous value of *a.minstart*, and computes the set of objects that are *attached* to *a*. The use of this function will become more evident in the next section, where we generate a demon that applied the conclusion of the rule to all pairs (*a,y*) produced by this derived term.

## 4.3 Code Generation

### 4.3.1 Generating Demons from Rules

When the compiler is applied to a set of rules, all rules are translated into an algebraic representation and all derivatives are computed. An algebraic rule is represented as (t => e), where t is the term from the relational algebra that is equivalent to the logical condition of the rule and e is the "conclusion" of the rule (the action part). A code fragment is then produced for each rule (t => e) and each derivative ∂t/∂R by applying the exp2 method as follows:

```
expF(∂T/∂R,x,y,u,v,e)
```

The result is a fragment of CLAIRE code that applies the conclusion of the rule to all pairs (u,v) of objects that satisfies the condition of the rule because of the event "x.R := y".

These fragments are combined, for each relation R, into a if-write demon that will be used to change the value of x.R, or R[x] if R is a table (in CLAIRE, tables and slots are abstracted into a binary relations and can be used indifferently in the logic language). The demon generated for R is based on the following pattern:

```
if_write[R](x, y)
    -> let LAST := get(R,x) in          // the previous value is stored
        (if (LAST != y)                 // checks that this is indeed an update
            (x.R := y,                  // performs the update
             …                          // propagation fragment from 1ˢᵗ rule
             if (x.R = y) …             // propagation fragment from 2ⁿᵈ rule
             if (x.R = y) …             // propagation fragment from 3ʳᵈ rule
             …))
```

Note that the propagation fragment obtained from the rules other than the first one are guarded with a test (x.R = y) to make sure that the rules are only applied if the update is still valid.

CLAIRE makes a difference between mono- and multi-valued relations. For multi-valued relations, the event model is simpler since the only event that is detected is the addition of a new member in one object's set of values. Thus the generation of the demon is slightly simpler in the multi-valued case.

For instance, the last phase of the compilation of the closure rule, which we gave as an example in the previous section, consists of associating the set of functions (derived from each relational term occurring in the condition) and the CLAIRE expression in the conclusion, so as to build one *if-write* demon per function. For instance, the demon associated to the rule closure and the relation edge is compiled as follows (it is evaluated each time a pair (x,y) is added to the edge relation):

```
if_write[edge](x, y)
    -> if (x.edge add? y)                // tests if y is not already here
       (x.path :add y,
        for z in y.path  x.path :add z)
```

The handling of instantiation is straightforward since each instance is attached to its class through an inversible relationship (isa/instances). Therefore the representation of the event

```
x :: C
```

is made with the term χ(minstart). On the other hand, generating a large collection of if-write demons for this "isa" relation would slow instantiation down for every class; thus, we associated directly the if-write demon with the class C (this is why C must be a constant in the expression x :: C).

### 4.3.2 Code Generation Techniques

The code generation phase is mostly the application of the expR and expF methods, which embodies the application of various optimisation schemes, such as the generation of simpler and faster code if a sub-term represents a mono-valued relation. Moreover, there are two other techniques that are used to produce procedural code, which is as close as possible to what the user would write herself to propagate the rules. The first technique is the use of precise type inference and the second one is the use of modes to distinguish between different semantics for applying the rules' conclusions.

Type Inference within the rule compiler is defined by associating a second-order type to each subterm, either from the relational algebra or the derived functional algebra. These second-order types are functions that infer an output type from any input type or reciprocally. For instance, if y is bound to x through the relation represented by t, this function will return the type of y depending on the type of x. The case of a derivative term is slightly more complex since there are two input type parameters, namely the type of x and y where the update is "x.R := y". These functions are obtained in a straightforward manner through the abstract interpretation of the relational operators on the CLAIRE type domain, which is an inclusion lattice [CC77]. This precise type inference is key to producing efficient code, and it is more powerful than the type inference performed by CLAIRE on the fragments generated by the rule compiler.

CLAIRE supports some additional tuning of its rules through the *mode* declaration. These modes are defined as follows, and enable the user to remedy some of the problems that occur with non-monotonic and non-commutative rules.

```
<modes> ≡ mode( default | once | set | <integer> )
```

Rules should not, in general, be written in such a way that the result is order-dependent. The order in which they are triggered depends on the events (the propagation pattern) and the order in which the rules were entered. If it becomes necessary to have a more precise control over this order, priorities may be used. A priority is an integer attached to the rule using the *mode* declaration. CLAIRE will ensure that rules with higher priority will be triggered first for each new event. Notice that there is no implicit stack structure for triggering rules: the events generated by the application of a first rule may cause a new rule to be evaluated before a second rule is applied to the original event.

The conclusion is applied to any pair of objects that is obtained through a logical derivation of the conclusion and the update. This assumes that the conclusion can be fired more than once when the logical expression is redundant (multiple derivation paths for the same pair). However, it may be wrong to apply the conclusion twice even if a pair is obtained from two different paths. Consider the following example:

```
strange1(x:Person) :: rule( x.age = 18 | x.age > 10   =>   give(x, $1000))
```

An update "John.age := 18" may cause the rule to be fired twice. The solution in CLAIRE is to use the mode(set) declaration before the rule, which will force the computation of the set of pairs before firing the conclusion (thus eliminating duplicates). Note that the use of an event pattern is precisely another way to get rid of this problem.

Similarly, rules should have a monotonous behavior, which means that their conclusion should not invalidate the condition. CLAIRE supports non-monotonous rules but the difficulties linked to the absence of a clear semantic remain. Let us consider a second example:

```
strange2(x:Person, y:Person) :: rule(
    x.age = 18 & y ∈ x.friends   =>   (x.age := 19, invite(x,y)))
```

Should the rule invite one friend (which one) or all friends when the age of John is set to 18? The default behavior as well as the "set" behavior will invite all friends. The mode(once) declaration will ensure that the conclusion is fired only once for each update event. Using non-monotonic rules is tricky, but it is sometimes very useful.

These modes are implemented as an additional layer in the generation of the code fragments for the demons. The priorities are used to order the code fragments. The "set" mode builds a temporary set of pairs to be updated before applying the conclusion. The "once" mode uses an exception handler to make sure that the conclusion is only fired once.

### 4.3.3 Application

Here are two examples of demons generated by CLAIRE from the rules that were proposed as examples. The first one is associated to the minStart slot and represents the two first rules. It is obtained by applying the code generation method *expF* to the two derived terms that we have shown in Section 3.3.

```
if_write[minStart](t, y)
      -> let LAST := t.minStart in              // the previous value is stored
           (if (LAST != y)                      // checks that this is indeed an update
               (t.minStart := y,                // performs the update

               // propagation fragment from 1st rule (R1)
               when Y := get(attached,t) in  Y.minStart :+ (y – LAST),

               // propagation fragment from 2nd rule (R3)
               if (t.minStart = y)
                  for I in taskInterval
                     if (y >= I.minStart & LAST < I.minStart & t.maxEnd <= I.maxEnd)
                        I.duration :+ x.duration )
```

The second example is obtained from rules RS1 and RS2 for the location attribute:

```
if_write[location](x, y)
      -> let LAST := x.location in              // the previous value is stored
           (if (LAST != y)                      // checks that this is indeed an update
               (x.location := y,                // performs the update

               // propagation fragment from 1st rule (RS1)
               for Y in of~(location,y)         // reads the inverse of location
                  if (Y.strength >= x.strength)   failure(x)

               // propagation fragment from 2nd rule (RS2)
               if (x.location = y)
                  if (y.forest? & not(LAST.forest?)) computeVisibility(x) )
```

These two examples illustrate our claim that the rule compiler is able to produce procedural code for propagation that is very close to what a user would write.

### 4.3.4 Performance Results

We have not tested the new features of ClaireRules from a performance point of view due to the absence of established benchmarks. On more classical problems, such as those reported in Table 3, the code generation strategy of CLAIRE is very efficient (between 100 K to 50 Mips = Million *inferences* per second), even when solving complex problems. . The CPU time is measured on a Pentium III laptop at 500Mhz, using MSVC++ 6.0. Since CLAIRE can also generate Java code, we also include the results using Symantec's Visual Café 3.0. CLAIRE is between 2 or 3 orders of magnitude faster than what we have measured with RETE-based inference engines [For82] and various successive improvements such as TREAT [Mi87] or GATOR [HH93]. We have tested over the last 10 years four major commercial products – they are all based on refinements of the RETE algorithm - and two academic inference engines. The problems that we use in Table 3 are taken from a small set of the rare examples that circulate in the inference engine community, they *do not* exercise other features than rule propagation – this is not a backtracking benchmark. More empirically, we have noticed that CLAIRE programs written with production rules yield only about a 10 to 50 % penalty compared with hand-optimized procedural code.

|  | CPU time<br>**Java version** | CPU time<br>C++ version | **Rules/s**<br>C++ version |
|---|---|---|---|
| Simple Filtering | 0.01s | 0.02s | 50 Mips |
| Planning (Monkey & Banana) | 0.03s | 0.015s | 13 Mips |

| Constraint Propagation (triplet) | 0.36s | 0.05s | 3 Mips |
| --- | --- | --- | --- |
| Rules & methods (Airline) | 0.05s | 0.01s | 100Kips |
| Problem solving (Zebra) | 0.043s | 0.011 | 2.7 Mips |
| Problem solving (Dinner) | 0.18s | 0.08s | 250Kips |

Table 3: *Benchmarks on rules.*

The demons are produced using CLAIRE, taking advantage of CLAIRE high-level of abstraction, which simplifies the handling and the iteration of set expressions, as well as the reflective nature of the CLAIRE system, which makes generating new programs much easier. The CLAIRE program is then compiled into C++ or Java. The generation of Java programs is a new feature from the CLAIRE compiler. Our next step is to generate directly Java beans that exploit the event/listener model. Because we can generate Java code with no overhead or central control algorithm, this technique is ideally suited for compiling a set of rules into a Java bean component [Eng97]. This approach can be applied to all other publish & subscribe message architecture, which makes this technique very interesting for Enterprise Application Integration (EAI). For instance, we plan to use this approach to implement distributed workflow engines where a set of process rules is compiled into a group of distributed agents.

## 5. Search Algorithms

### 5.1 Search in an Imperative Language

In addition to rules, we saw that CLAIRE also provides the ability to do some hypothetical reasoning. This possibility to store successive versions of the database and to come back to a previous one is called the world mechanism. The part of the database that supports these defeasible updates is defined by the programmer and may include slots, tables, global variables or specific data structures such as lists or arrays. When the data model is defined, one may specify which slots are defeasible and which are not. Each time we ask CLAIRE to create a new world, CLAIRE saves the status of defeasible slots (and tables, variables, ...). Worlds are represented with numbers, and creating a new one is done with choice(). Returning to the previous world is done with backtrack(). Returning to a previous world n is done with world=(n). Worlds are organized into a stack (one cannot explore two worlds at the same time) so that save/restore operations are very fast. The current world that is being used can be found with world?(), which returns an integer. In addition, one may accept the hypothetical changes that were made within a world while removing the world and keeping the changes. This is done with the commit() method. commit() decreases the world counter by one, while keeping the updates that were made in the current world. It can be seen as a collapse of the current world and the previous world. Note that although the wording is closer to AI concepts, the implementation of worlds using a stack is very close to the trailing stack of a constraint logic programming system such as CHIP[VH89] and totally similar to the backtracking stack of a constraint solving tools such as ILOG solver [Pu94].

Here is a real example of the search strategy for the jobshop problem [Lab98]. The functional parameter tightest is a heuristic that returns the set of tasks (an Interval) that share the same machine and that looks for the most difficult to process.

```
solve(tightest:property) : boolean
    -> when I:Interval := tightest() in
        exists(t in {x in I | canbefirst?(I, t)} |
            branch( (for t' in (I but t)  t.After :add t',  solve(tightest) )
        else true                                  // no interval was found
```

This example shows many interesting features. Some of them have already been mentioned: the iteration of a selection set (the set of tasks that can be the first in the set I), the use of the pattern (I but t) or the use of a functional parameter. We also notice the defeasible update t.After :add t', which adds t' to the list of tasks performed after t and triggers the propagation rules that represent the scheduling constraints [Lab98].

Another interesting aspect of the branch control structure is contradiction handling (failure in a sub-tree). CLAIRE supports exceptions (represented as objects in a hierarchy) with a built-in contradiction subclass. When a contradiction is raised (for instance, by rule propagation), it is caught by the branch(...) instruction which backtracks and returns false. Thus, branch support a concise and readable programming style of search algorithms, an asset for defining easily new variations (other search strategies, instrumentation, ...).

We are also making good use of the extensibility of the iteration compiler. A task interval I is a set of tasks that share a same resource I.Res whose time windows are included in a given interval [Lab98]. Each task has a unique index and the set of tasks that belong to I is represented by a bit-vector I.SET. Thus the iteration of a task interval consists of iterating the set of tasks on the right machine whose indexes belong to the bit-vector. For instance, the iteration in the previous algorithm will be expanded into:

```
           for t in (I.Res).users  (if I.SET[t.index]  (if canbefirst?(I, t)  branch(...) ))
```

The ability to pass function parameters at no cost from a performance point of view is a key feature to write parametric algorithms. This is reinforced by the fact that these parameters can be in-line methods, which means that the original in-line method can be seen as a higher-level function generator. For instance, we can use the following definition of max to define the method later:

```
// a generic max method
max(s:subtype[integer], greater:property, default:any) : integer
    => let x := default, empty := true in
          (for y in s  (if empty   (x := y, empty := false)
                        else if greater(x, y)  x := y),  x)
before(x:task, y:task) : boolean  =>  x.atleast ≤ y.atleast
later(x:Interval, t:task) : integer  ->  max(x but t, before, EndTask)
```

The code generated for the method later will be:

```
let t0 := EndTask, empty := true in
    (for t1 in (x.Res).users
        (if x.SET[t1.index]
           (if (t != t1)
              (if empty   (t0 := t1, empty := false)
               else if (t0.atleast ≤ t1.atleast)  t0 := t1))),  t0)
```

In that example, we have used max as a high-level code generator. This is the first step in writing truly reusable code for complex problem solving algorithms.

## 5.2 Composition Polymorphism

Multiple layers of rewriting rules lead to the need for a new paradigm to implement simplification (or reduction) rules. Consider the two following equations:

```
(A + B)[i,j] = A[i,j] + B[i,j]
x ∈ (s but t) = (x ∈ s) & (x != t)
```

These equations may be seen as simplification rules since they respectively allow us to extract a cell of a matrix sum without computing this sum explicitly or evaluate membership to a "but" set without actually computing this set. Usually, there is no need to represent these rules because the programmer uses them implicitly. However, this is no longer the case when these sequences of source-to-source rewriting are used. For instance, there may exist the expression M[i,j] in the body of an inline method that gets transformed into (A + B)[i,j] because of the form of the input parameters. This is even more obvious with the second example if we apply a set inline method to the set expression (s but t).

Representing these simplification rules leads to the notion of **composition polymorphism**. What we need to represent is that the processing of f(...,x,...) may be different when x is obtained as g(....). For instance, f is get (get is used implicitly in the form ...[...], e.g. M[i,j] = get(M,i,j)) and g is + in the first case; f is ∈ and g is but in the second case. This type of polymorphism might be captured with a complex abstract data type in a statically typed language using complex pattern matching rules, but is out of reach with our concrete type approach. This is why we have introduced the notion of pattern: a pattern is a set of function calls with a given selector and a list of types (list of types to which must belong the arguments to the call); a pattern in CLAIRE is written p[tuple(A,B,...)]. Patterns support composition polymorphism through the definition of new inline methods whose signature are made either of any (for concrete parameters) or patterns (for expressions). The compiler uses such a restriction when it finds a function call that matches its signature. Here is the application to the "matrix" example:

```
get(x:+[tuple(Matrix,Matrix)],i:any,j:any)
     => (args(x)[1][i,j] + args(x)[2][i,j])
```

Suppose that we now have this situation:

```
inverseTrace(M:Matrix) => sum(list{1.0 / M[i,i] | i in (1 .. N)})
f(M1:Matrix, M2:Matrix) -> inverseTrace(M1 + M2) > 0.0
```

The macro-expansion of inverseTrace(M1 + M2) will yield to an expression that involves (M1 + M2)[i,i]. This call matches the signature of the optimizing restriction that we defined previously (with x = M1 + M2 and args(x) = (M1,M2)). Thus a further reduction is performed and the result is:

```
let x := 0, i := 1 in
    (while (i >= N)
         (x :+ 1.0 / (M1[i,i] + M2[i,i]), i :+ 1),
      x > 0.0)
```

The macro-expansion of inverseTrace(M1 + M2) is performed by the compiler under the same considerations an any inlining optimization: the compiler will check that no side-effects are involved, or that the parameter of an inlining pattern is used exactly once.

## 5.3 Writing Elegant Algorithms

Our last example is taken from a real application that uses the Hungarian Algorithm to solve a bipartite matching problem (useful for global constraint propagation [Lab98]). Although it does not use search mechanisms, we believe that it demonstrates the value of CLAIRE high-level features to express a complex algorithm in a more readable and elegant manner, even if not declarative.

```
// builds a maximum weight complete matching
match() -> (...,                                                            // initialization
            while (HN != N)   (if not(grow())  dualChange()))               // main loop
// a step repeats until the forest is hungarian (return value is false)
// or the matching is improved (return value is true)
// explore is the stack of even nodes that have not been explored yet
grow() : boolean
   -> let i := pop(explore) in          // explore is the stack of nodes that need to be explored
         (exists(j in {j in GpiSet(i,LastExplored[i] + 1,LastValid[i]) | not(odd?[j])} |
               (if (sol-[j] != 0)
                    (//[SPEAK] grow: add (~S,~S) to forest// i,j,
                     odd?[j] := true, pushEven+(sol-[j]), tree[j] := i, false)
                 else (augment(i,j), true))) |
            (if not(empty?(explore))    // if we did not find a j to augment the forest,
                grow()) )               // grow is called recursively
// change the dual feasible solution, throw a contradiction if there are no perfect matching
dualChange() : integer
   -> let e := Min( list{vclose[i] | i in {j in Dom | not(odd?[j])}}) in
        (//[SPEAK] DUAL CHANGE: we pick epsilon = ~S // e,
         if (e = NMAX)  contradiction!(),
         for k in stack(even)  (pi+[k] :+ e, LastExplored[k] := LastValid[k]),
         for j in {j in Dom | odd?[j]}  (pi-[j] :- e, vclose[j] := NMAX)),
      clear(explore),
      for i in stack(even)
         let l := Gpi[i], k := size(l), toExplore := false in
            (while (LastValid[i] < k)
               (k, toExplore) := reduceStep(i,j,l,k,toexplore),
                if toExplore  push(explore,i)))
// look at edges outside the valid set one at a time
reduceStep(i:Dom,j:Dom,l:list,k:integer,toExplore:boolean) : tuple(integer,boolean)
   -> let j := l[k], c := Cpi(i,j) in
         (if (c = 0)
             (//[SPEAK] dual_change: Add edge ~S,~S // i,j,
              Gpiadd(l,i,j,k), toexplore := true)
          else (vclose[j] :min c, k :- 1)),
       list(k,toexplore))
```

Let's briefly comment the three functions above. match builds the matching by repeatedly applying the two steps grow and dualChange until the complete set of nodes is spanned (this is a primal-dual algorithm). grow builds a Hungarian forest using a classical search for augmenting chains (with alternate marks even and odd) in the reduced graph Gpi. dualChange changes the dual values stored in the arrays pi+ and pi- so that the reduced graph Gpi is augmented. The reduced graph is implicitly defined as the subset of all edges satisfying a given condition on the dual weights (that the corresponding constraint in the linear model is saturated). It is represented by the following pattern: GpiSet(i,a,b) refers to the subset of neighbors of the node i in this reduced graph where a and b are two iteration cursors (we maintain an ordered set of edges incident to i in Gpi; all edges up to position a have been explored and all edges up to position b have been touched).

A complete explication of the algorithm is out of the scope of this paper, but we can notice:

- The use of patterns such as stack(even) or GpiSet(i,a,b) which encapsulates the complexity due to data structure handling (e.g., using the even array to implement the stack).

- The use of complex set expressions that are iterated (e.g., the first line of `dualChange`) and the use of set operators (`exists`) and generic functions (`min`).
- The use of multi-valued functions (`reduceStep`) to isolate fragments of code that do not return a single value.
- The use of active comments, which can be used as trace statements.

From our point of view, a description of this complex algorithm in less than 30 lines is very valuable, and is a strong advocate for using high-level languages. The interest of the techniques presented in this paper is that they may come with no penalties for run-time performance, since this CLAIRE implementation was checked to be exactly as efficient as a standard C++ implementation. These 30 lines may be compared with the 60 lines of the C++ program that was written by hand or the 90 lines that were generated by the CLAIRE compiler. The real gain is not the concision, it is the ability to grasp the design of the algorithm at a higher level of abstraction while reviewing the code.

# 6. Conclusion

Experiments in the previous years have shown that constraint programming was best used in combination with other optimization techniques for combinatorial problem solving. The fact that hybrid algorithm engineering cannot be always performed exclusively through the use of components and may require some programming motivates the use of multi-paradigm programming languages. There have been many proposals for incorporating "advanced features" such as sets or rules into object-oriented languages, especially in the AI community. On the other hand, few have managed to incorporate as well a high degree of polymorphism and the use of higher-order functions. Among those few, the most interesting, from our point of view, are LIFE and Oz.

The major contribution of LIFE [AK93] is to provide with a clean integration of logic, inheritance and functions (as its name stands). LIFE played an important role when we tried to put some elegance into our "mix of features". However, there is a top-down vs. bottom-up difference in the design strategy of CLAIRE and LIFE. CLAIRE is the bottom-up integration of features that we know how to compile efficiently whereas LIFE is the top-down implementation of paradigms that are known to blend elegantly. The result is that it is easier, from our point of view, to write efficient algorithms with CLAIRE than with LIFE.

Oz is a concurrent constraint language [Smo95] belonging to the post-CLP language generation. Its object-oriented kernel offers multiple inheritance and a clean integration of higher order functions, in particular with the predefined search combinators that offer parametric search procedures [SS94]. These predefined combinators have been extended into the ability to define search procedures (programming constraint inference engines [Sch87]) with few lines of elegant code, which is very similar to CLAIRE and LAURE [Cas94], both from a motivation and a result point of view,

The comparison between Oz and CLAIRE is interesting because Oz is both more sophisticated and simpler than CLAIRE. On the one hand, Oz is more ambitious than CLAIRE since it proposes a computational model which is original, centered on constraints and search (stores), and supports distributed computing. CLAIRE uses an object-oriented data model that is geared towards knowledge representation, with the additional benefits of providing a strong basis for data types. Oz has been a model for some of CLAIRE features, such as the use of higher-order functions to capture parametric search strategies. On the other hand, Oz is still an un-typed language that is too high-level, from our point of view, to be a language of choice to implement propagation strategies. Oz relies on the ability to use a lower-level language to implement some of the constraint propagation strategies, while CLAIRE is a self-contained tool.

To summarize, we have presented a high-level language for writing re-usable algorithms. The application of this language to various optimization problems has resulted in programs that are easier to read (and thus to maintain) and to reuse (or to combine). This is mostly due to two features:
- Declarative programming is encouraged with the use of logic rules (the programmer is freed from the concern about propagation patterns), which can be compiled into procedural demons with no run-time processing overhead.
- Set programming and set iteration raise the level of abstraction and recursively-nested expressions can be compiled with no run-time processing overhead.

We believe that our experience with CLAIRE may be of interest to the CLP community on three topics: the technology of compilation-by-differentiation for rules, the definition of a simple multi-paradigm language supported by real applications and the concern for a general purpose programming language for hybrid algorithms. On this last point, CLAIRE offers different tools for different concerns: an object oriented type system for elegant modeling, a data-driven rule compiler for declarative programming, parametric polymorphism, iteration and patterns for data structure manipulations and a version mechanism (reified in SaLSA [LC98]) for controlling search trees. This suggests that the scope of CLAIRE is broader than its current use for combinatorial optimization algorithms and that it could be used for applications that have been associated with logic and rule-based programming.

CLAIRE has been used for various deployed industrial applications, in different companies. It was also used as a support tool of choice for teaching "Constraints and Algorithms". CLAIRE is written in CLAIRE (10,000 lines of code), except for a small bootstrapping, run-time kernel, and the optimizer makes heavy use of CLAIRE features (including garbage collection). CLAIRE is a public domain tool that may be downloaded from various WEB sites.

# References

[AK93]   H. Aït-Kaci. *An Introduction to LIFE: Programming with Logic, Inheritance, Functions and Equations*. Proceedings of ILPS'93, p.117, 1993.

[AS97]   K.R. Apt, A. Schaerf. *Search and Imperative Programming*. Proceedings of POPL'97, p. 67-79, ACM Press, 1997.

[Cas94]  Y. Caseau. *Constraint Satisfaction with an Object-Oriented Knowledge Representation Language*. Applied Intelligence, Vol. 4, no. 2, p. 157-184, May 1994.

[Cas95]  G. Castagna. *Covariance and Contravariance: Conflict without a Cause*. ACM Transactions on Programming Languages and Systems, Vol 17, no. 3, p. 431-447, May 1995.

[CL94]   C. Chambers, G. Leavens. *Typechecking and Modules for Multi-Methods*. Proceedings of OOPSLA'94, ACM Sigplan Notices, p. 805-843, Portland, 1994.

[CL96a]  Y. Caseau, F. Laburthe. *Introduction to the CLAIRE Programming Language*. LIENS research report 96-16, Ecole Normale Supérieure, 1996.

[CL96b]  Y. Caseau, F. Laburthe. *Cumulative Scheduling with Tasks Intervals*. Proceedings of JICSLP'96, M. Maher ed., p. 363-377, The MIT Press, 1996.

[CP93]   Y. Caseau, L. Perron. *Attaching Second-Order Types to Methods in an Object-Oriented Language*. Proceedings of ECOOP'93, O.M. Nierstrasz Ed., p.142-160, Kaiserslautern, Germany, 1993.

[DGL95]  M. Day, R. Gruber, B. Liskov, A. Myers. *Subtypes vs. Where Clauses: Constraining Parametric Polymorphism*. Proceedings of OOPSLA'95, ACM Sigplan Notices, p. 156-168, Austin, TX, 1995.

[Eng97]  R. Englander. *Developing Java Beans*. O'Reilly, 1997.

[For82]  C.L. Forgy. *Rete: A Fast Algorithm for the Many Pattern/Many Object Pattern Match Problem* Artificial Intelligence, 19, p. 17-37, 1982.

[GR83]   A. Goldberg, D. Robson. *Smalltalk-80: The Language and its Implementation*. Addison-Wesley, 1983.

[HH93]   E. Hanson, M. Hasan. *Gator: An Optimized Discrimination Network for Active Database Rule Condition Testing*. Tech report 93-036, CSE CIS Department University of Florida Gainesville, FL, 1993.

[HU94]   U. Holzle, D. Ungar. *A Third-Generation Self Implementation: Reconciling Responsiveness with Performance*. Proceedings of the ACM OOPSLA'94 Conference, p. 229-243, Portland, OR, October 1994.

[Lab98]  F. Laburthe. *Constraints and Algorithms in Combinatorial Optimization*, Ph.D. Thesis (in French) University of Paris VII, 1998.

[LC98]   F. Laburthe, Y. Caseau. *SaLSA: A Language for Search Algorithms*, Proceedings of CP'98, M.Maher, J.-F. Puget eds., Springer, LNCS 1520, p.310-324, 1998.

[KW89]   M. Kifer, J. Wu. *A logic for Object-Oriented Logic Programming (Maier's O-Logic Revisited)*. Proceeding of PODS-89, Philadelphia, 1989.

[McL81]  B.J. MacLennan. *Programming with A Relational Calculus*. Rep N° NPS52-81-013, Naval Postgraduate School, September 1981.

[Mir87]  D.P. Miranker. *TREAT: A better match algorithm for AI production systems*. In Proceedings AAAI-87 Sixth National Conference on Artificial Intelligence, pages 42-47. Morgan Kaufmann, San Francisco, August 1987.

[NT89]   S. Naqvi, S. Tsur. *A Logical Language for Data and Knowledge Bases*. Computer Science Press, 1989.

[Pug94]  J.-F. Puget. *A C++ Implementation of CLP*. Ilog Solver Collected papers, Ilog tech. report, 1994.

[Sch87]  C. Schulte. *Programming Constraint Inference Engines*. Proceedings of CP97, G. Smolka ed., Springer, LNCS 1330, p. 519-533, 1997.

[SDD86]  J. Schwartz, R. Dewar, E. Dubinsky, E. Schonberg. *Programming with Sets: an Introduction to SETL*. Springer, New-York, 1986.

[SKG87]  H. Schmidt, W. Kiessling, V. Guntzer, R. Bayer. *Compiling Exploratory And Goal-Directed Deduction into Sloppy Delta-Iteration*. Proc. of the Symposium on Logic Programming, p. 233-243, San Francisco, 1987.

[SS94]   C. Schulte, G. Smolka. *Encapsulated Search for Higher-order Concurrent Constraint Programming*, Proceedings of ILPS'94, M. Bruynooghe ed., MIT Press, p. 505-520, 1994.

| | | |
|---|---|---|
| [Smo95] | G. Smolka. *The Oz Programming Model*, Proceedings of Computer Science Today, J. Van Leeuwen ed., LNCS 1000, p. 324-343, Springer, 1995. | |
| [SZ88] | D. Sacca, C. Zaniolo. *Differential Fixpoint Methods and Stratification of Logic Programs*. MCC Technical Report ACA-ST-032-88, January 1986. | |
| [VH89] | P. Van Hentenryck. *Constraint Satisfaction in Logic Programming*. The MIT press, Cambridge, 1989. | |

## Appendix: Benchmark Results

The CLAIRE language is not a powerful high-level language whose compiler has progressively been designed to eliminate undue overhead. On the contrary, CLAIRE was a simple C++ code generator, which was designed to guarantee the absence of overhead, and that has evolved into a powerful language over the years while we were improving the compiler technology. The first application of CLAIRE, in 1994, has been job shop scheduling software; we presented a new state-of-the-art algorithm with a new data structures (task intervals). Because we did not have the techniques presented here available at that time, the handling of the data structures was done by hand everywhere in the code. An encapsulation approach based on iterators would have produced a slow-down of a factor of 10, but we were able to rewrite the algorithm two years later with exactly no overhead, since the compiler had been tuned to produce precisely the code we were starting from. We have used CLAIRE to encode many classical graph algorithms (shortest path, flows, matching, etc.). With earlier versions of the compiler, the encoding was very similar to the C or Pascal algorithms that we were copying from. The introduction of the techniques presented here allowed us to evolve towards much more elegant and compact encoding, without any overhead.

This paper contains three sets of benchmarks that we use to monitor CLAIRE performance; the first two are reported in this appendix and the last one was shown in Section 4.3.4. They illustrate our central claim that CLAIRE introduces no overhead compared to more traditional procedural languages. We first gathered two "algorithmic" micro-benchmark suites: respectively the Self and the Stanford benchmarks. Self's original results [HU94] are not reproduced since the hardware has changed too much in the last few years; but the comparison with C++ shows that CLAIRE seems faster than Self.

| | | CLAIRE | Visual C++ 6 | Java (Visual Café 3) |
|---|---|---|---|---|
| | Fibonacci(38) | 4,140 | 4,090 | 4,530 |
| **Self benchmarks** | Fill a vector of 100,000 integers (x 1,000) | 1,220 | 1,190 | 1,200 |
| | Increment each element of that vector (x 1,000) | 1,800 | 1,750 | 1,830 |
| | Add up all the elements of that vector (x 1,000) | 830 | 890 | 970 |
| | Sum iterators of 2 nested 100-loops (x 100,000) | 6,970 | 6,940 | 6,920 |
| **Stanford benchmarks** | Bubble sort of 10,000 integers | 890 | 840 | 1,140 |
| | Quick sort of 2,000,000 integers | 620 | 560 | 760 |
| | Tree sort of 5,000 integers | 660 | 780 | 1,730 |
| | Multiplication of 2 300x300-matrices of integers | 500 | 520 | 1,190 |
| | Multiplication of 2 300x300-matrices of floats | 2,080 | 1,540 | 2,420 |
| | Permutation of the elements of a vector | 3,090 | 2,520 | 2,730 |
| | Time-consuming placement problem (x 500) | 7,610 | 6,660 | 7,470 |

We also report benchmark results on libraries for collection data structures for the same 3 languages (we chose the closest C++ and Java libraries to CLAIRE lists and sets). Moreover, multiple experiments have confirmed what these benchmarks suggest: handling collection data structure is definitely a strong feature of CLAIRE.

|  |  | CLAIRE | Visual C++ 6 | Java (Visual Café 3) |
|---|---|---|---|---|
| **Lists of 1,000 objects** | Creation | 140 | 90 | 390 |
|  | Iteration | 2,690 | 7,260 | 26,700 |
|  | Membership | 4,580 | 15,600 | 17,800 |
|  | Deletion and insertion | 1,840 | 5,770 | 6,970 |
| **Sets of 1,000 objects** | Creation | 340 | 580 | 6,090 |
|  | Iteration | 2,440 | 7,220 | 26,200 |
|  | Membership | 300 | 750 | 133,000 |
|  | Deletion and insertion | 4,120 | 4,950 | 2,450 |
| **Hash sets of 1,000 objects** | Creation | 170 | 8,770 | 1,730 |
|  | Iteration | 3,750 | 23,800 | 29,900 |
|  | Membership | 90 | 8,030 | 370 |
|  | Deletion and insertion | 60 | 3,170 | 870 |